\documentclass[11pt]{article}
\usepackage[margin=1in]{geometry}
\usepackage[T1]{fontenc}
\usepackage[utf8]{inputenc}
\usepackage{lmodern}
\usepackage{amsmath,amssymb}
\usepackage{booktabs,longtable,array}
\usepackage{hyperref}
\usepackage{xurl}
\usepackage{cleveref}
\usepackage[most]{tcolorbox}
\usepackage{xcolor}
\usepackage{enumitem}
\usepackage{booktabs}
\usepackage{fancyvrb}
\usepackage{tikz-cd}
\usepackage{listings}
\usepackage{fontawesome5}
\usepackage{placeins}

\hypersetup{
  linkcolor=blue,
  colorlinks=true,
  urlcolor=blue,
  citecolor=blue,
  pdfborder={0 0 1}
}

\definecolor{shellbg}{RGB}{245,245,245}
\definecolor{shellframe}{RGB}{215,215,215}
\definecolor{pfInk}{HTML}{1F2937}
\definecolor{pfMuted}{HTML}{6B7280}
\definecolor{inputBg}{HTML}{EEF4FF}
\definecolor{inputBorder}{HTML}{C7D7FE}
\definecolor{filterBg}{HTML}{ECFDF5}
\definecolor{filterBorder}{HTML}{A7F3D0}
\definecolor{chipBg}{HTML}{FFF1F2}
\definecolor{chipBorder}{HTML}{FDA4AF}
\definecolor{chipText}{HTML}{BE123C}
\definecolor{opfAccount}{HTML}{7C3AED}
\definecolor{opfAddress}{HTML}{1D4ED8}
\definecolor{opfDate}{HTML}{EA580C}
\definecolor{opfEmail}{HTML}{0F766E}
\definecolor{opfPerson}{HTML}{BE123C}
\definecolor{opfPhone}{HTML}{15803D}
\definecolor{opfUrl}{HTML}{0369A1}
\definecolor{opfSecret}{HTML}{B45309}

\newcommand{\colorlabel}[2]{\textcolor{#1}{#2}}

\newcommand{\opfcolorlegend}{%
  \colorlabel{opfAccount}{account\_number} | %
  \colorlabel{opfAddress}{private\_address} | %
  \colorlabel{opfDate}{private\_date} | %
  \colorlabel{opfEmail}{private\_email} | %
  \colorlabel{opfPerson}{private\_person} | %
  \colorlabel{opfPhone}{private\_phone} | %
  \colorlabel{opfUrl}{private\_url} | %
  \colorlabel{opfSecret}{secret}%
}

\newcommand{\piichip}[1]{%
  \tcbox[
    on line,
    enhanced,
    nobeforeafter,
    box align=base,
    colback=chipBg,
    colframe=chipBorder,
    coltext=chipText,
    boxrule=0.6pt,
    arc=0pt,
    left=0pt,
    right=0pt,
    top=0pt,
    bottom=0pt,
  ]{\sffamily\scriptsize #1}%
}

\newtcblisting{shellsample}[1][]{%
  enhanced,
  breakable,
  colback=shellbg,
  colframe=shellframe,
  boxrule=0.4pt,
  arc=2pt,
  left=8pt,
  right=8pt,
  top=6pt,
  bottom=6pt,
  title=\texttt{#1},
  fonttitle=\small\ttfamily,
  coltitle=black,
  colbacktitle=gray!18,
  listing only,
  listing options={
    basicstyle=\ttfamily\small,
    breaklines=true,
    breakatwhitespace=false,
    columns=fullflexible,
    keepspaces=true,
    showstringspaces=false,
    upquote=true,
  },
  before skip=0.75\baselineskip,
  after skip=0.75\baselineskip,
}

\title{Model Card for OpenAI Privacy Filter}
\author{
{\small Charles de Bourcy$^*$, Sahra Ghalebikesabi$^*$, Avi Schwarzschild$^*$, Alex Gorbachev$^*$, Mihai Maruseac$^*$} \\
{\small Annie Chu, Vol Kyrylov, Tong Mu, Ally Bennett, Andy Nguyen, Casey Meehan, Jessica Gan Lee} \\
{\small Shane Bauer, Harold Nguyen, Rodolpho Eckhardt, Yuqi Liu, Charlie Oxborough, Marco Rougeth} \\
{\small Omar Chedid, Caio Costa, Yash Parikh, Yao Li, Congzheng Song, Om Thakkar, Vinnie Monaco} \\[0.35em]
{\footnotesize $^*$Co-lead authors}
}
\date{\today}
\begin{document}

\maketitle

\begin{abstract}
OpenAI Privacy Filter is a compact, bidirectional token-classification model for detecting and redacting personally identifiable information (PII) and secrets in unstructured text. 
The model is derived from an autoregressively pretrained checkpoint and converted into a bidirectional, banded-attention classifier that labels an input sequence in a single forward pass.
A constrained Viterbi decoder produces coherent spans across eight privacy categories and exposes configurable operating points for precision--recall tradeoffs. 
Privacy Filter has 1.5 billion total parameters, 50 million active parameters per token, and a 128,000-token context window. 
It is designed for efficient local deployment and domain-specific fine-tuning. 
Privacy Filter is intended as a configurable data-minimization component within layered privacy workflows, not as an anonymization or compliance guarantee.
\end{abstract}

\pagebreak
\tableofcontents
\pagebreak

\section{Introduction}
\label{sec:introduction}

OpenAI Privacy Filter is a bidirectional token-classification model for personally identifiable information (PII) detection and redaction in text. It is designed for high-throughput privacy workflows, and is able to perform context-aware detection of PII in unstructured text.

\begin{figure}[!htbp]
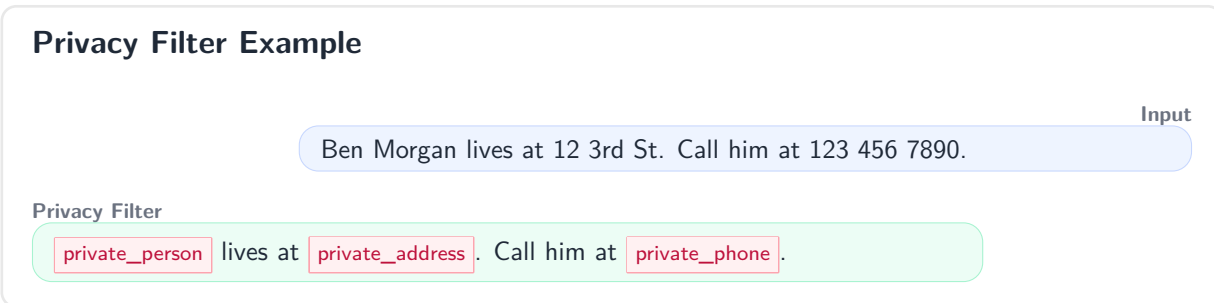

\centering
\begin{tcolorbox}[
  enhanced,
  width=0.98\linewidth,
  colback=white,
  colframe=gray!18,
  boxrule=1pt,
  arc=5pt,
  left=8pt,
  right=8pt,
  top=6pt,
  bottom=6pt,
]

{\sffamily\large\bfseries\color{pfInk} Privacy Filter Example}

\begin{flushright}
{\sffamily\scriptsize\bfseries\color{pfMuted} Input}
\begin{tcolorbox}[
  enhanced,
  width=0.77\linewidth,
  colback=inputBg,
  colframe=inputBorder,
  boxrule=0.4pt,
  arc=7pt,
  left=5pt,
  right=0pt,
  top=2pt,
  bottom=0pt,
  before skip=0pt,
  after skip=0pt,
]
\raggedright
{\sffamily\small\color{pfInk}
Ben Morgan lives at 12 3rd St. Call him at 123 456 7890.}
\end{tcolorbox}
\end{flushright}

\begin{flushleft}
{\sffamily\scriptsize\bfseries\color{pfMuted} Privacy Filter}
\begin{tcolorbox}[
  enhanced,
  width=0.82\linewidth,
  colback=filterBg,
  colframe=filterBorder,
  boxrule=0.4pt,
  arc=7pt,
  left=5pt,
  right=5pt,
  top=2pt,
  bottom=0pt,
  before skip=0pt,
  after skip=0pt,
]
{\sffamily\small\color{pfInk}
\piichip{private\_person} lives at
\piichip{private\_address}. Call him at
\piichip{private\_phone}.}
\end{tcolorbox}
\end{flushleft}

\end{tcolorbox}

\caption{Illustrative chat-style interaction showing Privacy Filter redacting PII from raw user text.}
\label{fig:chat-redaction-example}
\end{figure}

Privacy Filter is pretrained autoregressively to arrive at a checkpoint with similar architecture to gpt-oss, albeit of a smaller size. We then converted that checkpoint into a bidirectional token classifier over a privacy label taxonomy and post-trained it with a supervised classification loss. For architecture details about gpt-oss, see the \href{https://openai.com/index/gpt-oss-model-card/}{gpt-oss model card}. Instead of generating text token by token, Privacy Filter labels an input sequence in a single forward pass, then decodes coherent spans with a constrained Viterbi procedure. For each input token, the model predicts a probability distribution over the label taxonomy, which consists of the eight output categories detailed in \Cref{sec:model-architecture}. Privacy Filter provides the community with a free-to-download, easy-to-use, and tunable model with the following practical strengths:

\begin{itemize}[leftmargin=*]
  \item Permissive license:  We are releasing Privacy Filter under the Apache 2.0 license, allowing for experimentation, customization, and commercial deployment.
  \item Small footprint: Privacy Filter is a small model that can run in a web browser or on a laptop; it contains 1.5 billion total parameters and 50 million active parameters.
  \item Fine-tunable: While the out-of-the-box performance is strong, Privacy Filter can be further adapted to specific data distributions through simple and data-efficient fine-tuning.
  \item Long-context: The 128,000 token context window enables processing long text with high throughput and no chunking.
  \item Runtime control: Privacy Filter ships with span decoding that users can configure to control the precision-recall tradeoffs and detected span lengths.
\end{itemize}

Privacy protection for AI systems is an ongoing effort across research, product design, evaluation, and deployment.
Privacy Filter reflects one direction we believe is important: small, efficient models with frontier capability in narrowly defined tasks that matter for real-world AI systems. We are releasing it because we think privacy-preserving infrastructure should be easier to inspect, run, adapt, and improve.
Our goal is for models to learn about the world, not about private individuals. Privacy Filter helps make that possible.
We're releasing this preview of Privacy Filter to receive feedback from the research and privacy communities and iterate further on model performance.

\section{Model Details}
\label{sec:model-details}

Privacy Filter is a bidirectional token classification model with span decoding. It is trained in phases, beginning with autoregressive pretraining. The pretrained language model is then modified and post-trained as a bidirectional banded attention token classifier with band size 128 (effective attention window: 257 tokens including self). At inference time, we apply constrained sequence decoding to produce coherent \href{https://en.wikipedia.org/wiki/Inside\%E2\%80\%93outside\%E2\%80\%93beginning_(tagging)}{BIOES} (Begin, Inside, Outside, End, Single) span labels.

\subsection{Model Architecture}
\label{sec:model-architecture}

As a bidirectional attention-based model, Privacy Filter has several standard components. Privacy Filter uses a pre-norm transformer encoder-style stack. The model begins with token embeddings and then passes those representations through eight repeated transformer blocks. Within each block, attention uses grouped-query attention with rotary positional embeddings, with 14 query heads and 2 key-value heads, corresponding to a group size of 7 query heads per key-value head. The feed-forward sublayers are implemented as sparse mixture-of-experts blocks with 128 experts total and top-4 routing per token. The final layer is a token-classification head over privacy labels, rather than a natural-language vocabulary, and uses a residual-stream width of $d_{\mathrm{model}} = 640$.

Relative to iterative autoregressive approaches, this design allows all tokens to be labeled in one pass, which is important for throughput-sensitive applications. Relative to classical masked-language-model pretraining approaches, this is a post-training conversion of an autoregressive model rather than a native masked-LM setup.

\subsection{Output Shape}
\label{sec:output-shape}

Privacy Filter can detect 8 privacy span categories:
\begin{enumerate}[leftmargin=*, itemsep=0pt, topsep=2pt, parsep=0pt, partopsep=0pt]
  \item \texttt{account\_number}
  \item \texttt{private\_address}
  \item \texttt{private\_email}
  \item \texttt{private\_person}
  \item \texttt{private\_phone}
  \item \texttt{private\_url}
  \item \texttt{private\_date}
  \item \texttt{secret}
\end{enumerate}

To perform token-classification, each non-background span category is expanded into boundary-tagged token classes: \texttt{B-<label>}, \texttt{I-<label>}, \texttt{E-<label>}, \texttt{S-<label>}, plus the background class, \texttt{O}. So the total number of token-level output classes is 33: 1~background class + 8~span labels * 4~boundary tags = 33~classes. Thus, the output head emits 33 logits for each token. For a sequence of length $T$, the output has shape $[T, 33]$; for a batch of size B, it has shape $[B, T, 33]$.

The token-label vocabulary consists of the background label \texttt{O} plus BIOES-tagged variants of each privacy category: \texttt{account\_number}, \texttt{private\_address}, \texttt{private\_email}, \texttt{private\_person}, \texttt{private\_phone}, \texttt{private\_url}, \texttt{private\_date}, and \texttt{secret}. In other words, for each category, the model predicts \texttt{B-}, \texttt{I-}, \texttt{E-}, and \texttt{S-} forms corresponding to begin, inside, end, and single-token spans. At inference time, these per-token logits are decoded into coherent BIOES span labels using constrained sequence decoding.

\subsection{Sequence Decoding}
\label{sec:sequence-decoding-rationale-and-operating-point-calibration}

\subsubsection{Rationale}
\label{sec:rationale}

After the token classifier produces per-token logits, we decode labels with a constrained Viterbi decoder using linear-chain transition scoring, rather than taking an independent argmax for each token. The decoder enforces allowed BIOES boundary transitions and scores complete label paths with start, transition, and end terms, plus six transition-bias parameters that control background persistence, span entry, span continuation, span closure, and boundary-to-boundary handoff. This global path optimization is intended to improve span coherence and boundary stability by making each token decision depend on sequence-level structure, not just local logits, especially in noisy or mixed-format text where local token decisions alone can produce fragmented or inconsistent boundaries.

\subsubsection{Operating-Point Calibration}
\label{sec:operating-point-calibration}

Sequence decoding parameters can discourage long spans of background classifications while encouraging redaction span starts and continuations, yielding broader and more contiguous redaction for improved recall. They can also be configured the other way for improved precision. At runtime, users can tune parameters that control this tradeoff.

\section{Uses}
\label{sec:uses}

\subsection{Intended Uses}
\label{sec:intended-uses}

Privacy Filter is intended for PII span detection in raw text and for passing detected spans to downstream redaction workflows. A useful way to think about the model is as a data minimization component: it identifies candidate sensitive spans and their classes, and the surrounding system decides whether to mask, remove, replace, pseudonymize, alert, or route them for review.

Representative applications include:

\begin{itemize}[leftmargin=*]
  \item Sanitizing text at ingestion time before storage, indexing, analytics, or retrieval.
  \item Preparing corpora for model training, evaluation, logging, or internal sharing.
  \item Redacting personal data and secrets in enterprise documents, email, support transcripts, and application logs.
  \item Adapting the model to domain-specific privacy policies through fine-tuning and operating-point calibration.
\end{itemize}

\subsection{Out-of-Scope and Misuse}
\label{sec:out-of-scope-and-misuse}

Privacy Filter should not be treated as an anonymization, compliance, or safety guarantee, a substitute for policy review in high-stakes deployments, a universal privacy oracle with fixed behavior across all text genres and regions, or a legal determination system.

Potential misuse includes treating unredacted output as safe for release, relying on default operating points without validating them on the target distribution, or skipping domain adaptation when local policy requires materially stricter or materially broader redaction criteria.

\section{Bias, Risks, and Limitations}
\label{sec:bias-risks-and-limitations}

\subsection{Over-reliance Risk}
\label{sec:risk-over-reliance}

Privacy Filter is a redaction and data minimization aid, not an anonymization, compliance, or a safety guarantee. Over-reliance on the tool as a blanket anonymization claim would risk missing desired privacy objectives. Privacy Filter is best used as one of multiple layers in a holistic end-to-end privacy-by-design approach.

\subsection{Static Label Policy}
\label{sec:limitation-static-label-policy}

The model will only identify PII spans that match the trained label taxonomy and definitions. Real-life privacy use cases are varied and complex, and definitions of appropriate label policies and decision boundaries can differ. Thus model defaults may not satisfy organization-specific governance requirements without calibration/fine-tuning.

Privacy Filter does not support configuring label policies dynamically at runtime; instead, changing policies requires further fine-tuning of the model. The native label set and associated decision boundaries may not be appropriate for every use case. For example, the model's training policy aims to prioritize personal identifiers, often preserving context that is not strongly person-linked by design; some users may want to adjust this choice.

Performance may drop on non-English text, non-Latin scripts, or naming patterns or domains that are out of distribution compared to model training.

\subsection{Failure Modes}
\label{sec:failure-modes}

Like all models, Privacy Filter can make mistakes, such as: under-detection of uncommon personal names, regional naming conventions, initials, honorific-heavy references, or domain-specific identifiers; over-redaction of public entities, locations, or common nouns when local context is ambiguous; fragmented or shifted span boundaries in mixed-format text, long documents, or text with heavy punctuation and layout artifacts; missed secrets for novel credential formats, project-specific token patterns, or secrets split across surrounding syntax; and over-redaction of benign high-entropy strings, placeholders, hashes, sample credentials, or synthetic examples that resemble secrets.

These limitations can interact with demographic, regional, and domain variation. For example, names and identifiers that are underrepresented in training data, or that follow conventions different from the dominant training distribution, may be more likely to be missed or inconsistently bounded.

\subsection{High-Risk Deployment Caution}
\label{sec:high-risk-deployment-caution}

Additional caution is warranted in high-sensitivity settings such as medical, legal, financial, human resources, education, and government workflows. In these settings, both false negatives and false positives can be costly: missed spans may expose sensitive information, while excess redaction can remove material context needed for review, auditing, or downstream decision-making.

\subsection{Recommendations}
\label{sec:recommendations}

We recommend using Privacy Filter as part of a holistic privacy-by-design approach rather than as the basis for a blanket anonymization claim. Before production use, it's best to evaluate the model in-domain against local policy references. Task-specific fine-tuning should be used when local policy differs from the base decision boundaries. High-sensitivity workflows should also retain human review paths.

\section{How to Get Started with the Model}
\label{sec:how-to-get-started-with-the-model}

Privacy Filter can be used in multiple ways supported by the \texttt{opf} CLI tool in the repository. The CLI can be used directly to predict PII spans:

\begin{shellsample}[shell]
  $ opf "Ben Morgan lives at 12 3rd St. Call him at 123 456 7890."
  <PRIVATE_PERSON> lives at <PRIVATE_ADDRESS>. Call him at <PRIVATE_PHONE>.
\end{shellsample}

Span predictions from a file are obtained using the -f flag:

\begin{shellsample}[shell]
  $ opf -f text_file
\end{shellsample}

A more verbose output can be specified with the --format json flag:

\begin{shellsample}[shell]
  $ opf "Ben Morgan lives at 12 3rd St." --format json
\end{shellsample}

The output begins with a short summary:

\begin{shellsample}[output]
  summary: output_mode=typed spans=2 by_label=private_address:1, private_person:1 latency_ms=65.1 decoded_mismatch=no
\end{shellsample}

And it contains a JSON document that describes all the identified spans:

\begin{shellsample}[shell]
  {
    "schema_version": 1,
    "summary": {
      "output_mode": "typed",
      "span_count": 2,
      "by_label": {
        "private_address": 1,
        "private_person": 1
      },
      "decoded_mismatch": false
    },
    "text": "Ben Morgan lives at 12 3rd St.",
    "detected_spans": [
      {
        "label": "private_person",
        "start": 0,
        "end": 10,
        "text": "Ben Morgan",
        "placeholder": "<PRIVATE_PERSON>"
      },
      {
        "label": "private_address",
        "start": 20,
        "end": 31,
        "text": "12 3rd St.",
        "placeholder": "<PRIVATE_ADDRESS>"
      }
    ],
    "redacted_text": "<PRIVATE_PERSON> lives at <PRIVATE_ADDRESS>"
  }
\end{shellsample}

Finally, the output also contains a color coded version of the original text where each PII span uses a color matching its type:

\begin{tcolorbox}[
  enhanced,
  breakable,
  colback=shellbg,
  colframe=shellframe,
  boxrule=0.4pt,
  arc=2pt,
  left=8pt,
  right=8pt,
  top=6pt,
  bottom=6pt,
  title=\texttt{shell},
  fonttitle=\small\ttfamily,
  coltitle=black,
  colbacktitle=gray!18,
  before skip=0.75\baselineskip,
  after skip=0.75\baselineskip,
]
{\ttfamily\small
\textbf{color legend:}\\[2pt]
\opfcolorlegend\\[6pt]
\textbf{color coded text:}\\[2pt]
\textcolor{opfPerson}{Ben Morgan} lives at \textcolor{opfAddress}{12 3rd St.}
}
\end{tcolorbox}

Alternatively, the CLI may be used to redact spans by collapsing all categories to a single label:

\begin{shellsample}[shell]
  $ opf "Ben Morgan lives at 12 3rd St." --output-mode redacted
  <REDACTED> lives at <REDACTED>.
\end{shellsample}

If the default decoder operating point is not desired, the Viterbi arguments can be changed. Consult the \textbf{Decode} section of the \texttt{opf redact --help} screen.

Finally, for fine-tuning use \texttt{opf train} mode:

\begin{shellsample}[shell]
  $ opf train --output-dir finetuned/ dataset.jsonl
\end{shellsample}

Consult \texttt{opf train --help} for more information on fine-tuning and \texttt{opf --help} for more information about other options that the \texttt{opf} CLI tool supports.

\section{Training Details}
\label{sec:training-details}

Privacy Filter training is intended to provide frontier-level PII detection capabilities and targeted privacy-patterns in a small form factor.

It was trained on a mix of publicly available data and internally generated synthetic datasets, intended to cover both realistic natural text and targeted privacy-pattern diversity, and to refine the model's understanding of real vs. fake secrets in the context of software development.

In cases where ground truths were missing for publicly available data, we used a frontier model in the GPT-5 family for annotation with a 2x2 protocol: two prompt formats (structured JSON with explicit offsets and inline span tagging) crossed with two reasoning settings (medium and high).

Synthetic privacy datasets were constructed from public datasets by applying format-matching augmentation to increase subtype and surface-form diversity, inserting the resulting spans into a natural context, and running automated quality controls that removed examples with missing target spans, extraneous spans from the same taxonomy, or formatting failures.

\subsection{Label Taxonomy}
\label{sec:label-taxonomy-notes}

The available labels shipped with the model (prior to any customer fine-tuning) are listed below with short descriptions of their coverage.

\begin{itemize}[leftmargin=*, itemsep=1pt, topsep=2pt, parsep=0pt, partopsep=0pt]
  \item \texttt{private\_person}: The name of a private person, including usernames and handles that identify a specific person.
  \item \texttt{account\_number}: A credit card number, bank account number, or other account identifier.
  \item \texttt{private\_url}: A web URL or IP address that is meant for a private audience or identifies a private person.
  \item \texttt{private\_email}: An email address used for personal communication or that identifies a private person.
  \item \texttt{private\_phone}: A phone number associated with a private person.
  \item \texttt{private\_address}: A specific location or address associated with a private person.
  \item \texttt{secret}: An API key, password, or other credential.
  \item \texttt{private\_date}: The date of birth, birth year, or other datetime that identifies a private person.
\end{itemize}

These labels are broad categories and are intended to cover attributes that identify real people. As such, placeholders are not intended to be classified. For example, the \texttt{secret} class does not include example API keys or placeholder values.

\subsection{Training Procedure}
\label{sec:training-procedure}

Training Privacy Filter was done in stages. First, the base model is pretrained as a generative language model. Then, we modified the architecture slightly and post-trained with a supervised token classificaiton loss. Pretraining was done following the routines described for pretraining the gpt-oss models (\href{https://arxiv.org/pdf/2508.10925}{gpt-oss-120b \& gpt-oss-20b Model Card}).

The supervised PII classification training began from the autoregressive checkpoint described above, rather than from a randomly initialized encoder. First, we convert that pretrained model into a token classifier architecture by replacing the language-model output layer with a token-classification head over the privacy label taxonomy. At the same time, we relax the original causal attention mask into a bidirectional banded attention pattern so the model can use both left and right context while preserving the local-attention structure used at inference time. 

After this architectural conversion, the model was post-trained with a supervised token-level objective over BIOES-style labels, allowing it to learn span boundaries and category assignments directly from annotated examples. This phase of training is where the model is finally optimized directly for the downstream task of labeling each token in the input in one forward pass.

\subsubsection{Decoding and Calibration}
\label{sec:decoding-and-calibration}

Since the post-training for classification is over BIOES tokens, the final test time use is best done by respecting those span boundaries. This is not accomplished with a naive argmax style output. Therefore, the final aspect of training includes the decoding and operating-point selection procedure. Rather than reading off token labels independently, we apply constrained Viterbi decoding with linear-chain scoring to produce coherent span predictions and to enforce valid BIOES transition patterns. We then calibrated decoding behavior by measuring span-level precision, recall, F1, and F2 across held-out data, which let us choose operating points that reflect different application preferences for conservatism or coverage and lets end users tweak this easily.

\section{Evaluation}
\label{sec:evaluation}

\subsection{Evaluation Goals}
\label{sec:evaluation-goals}

Our evaluation is designed to measure how well Privacy Filter performs across a range of realistic detection settings. In particular, we assess detection quality across diverse text regimes, including variation in domain, format, and language conditions. We also evaluate sensitivity to label-policy definitions, since privacy-detection performance depends not only on whether a span is identified, but also on how the underlying label taxonomy is defined. Finally, we measure model accuracy on standard external benchmarks in order to characterize performance on established reference tasks and better understand strengths and weaknesses across evaluation settings.

\subsection{Evaluation Datasets}
\label{sec:evaluation-datasets}

We evaluate Privacy Filter on three datasets chosen to cover complementary privacy-detection settings. First, \href{https://huggingface.co/datasets/ai4privacy/pii-masking-300k}{PII-Masking-300k} is a large multilingual synthetic benchmark that supports broad measurement across privacy categories, languages, and textual formats. Second, \href{https://github.com/Samsung/CredData/tree/main}{CredData} consists of codebase text containing credential-like strings and is used to assess the model's ability to detect secrets in software and technical content. Third, \href{https://huggingface.co/datasets/mks-logic/SPY}{SPY} is a synthetic dataset of medical consultations and legal questions that provides a more domain-specific setting for evaluating privacy detection in sensitive, high-context text. Taken together, these datasets help us assess model behavior across general multilingual text, software-security data, and specialized professional domains.

\subsubsection{PII-Masking-300k}
\label{sec:pii-masking-300k}

We map labeled spans in the PII-Masking-300k dataset to the categories supported by Privacy Filter, excluding PII-Masking-300k categories that don't have a corresponding definition in Privacy Filter:

\begin{itemize}[leftmargin=*]
  \item \texttt{account}, \texttt{banknum}, \texttt{bic}, \texttt{creditcard}, \texttt{cryptoaddress}, \texttt{docnum}, \texttt{driverlicense}, \texttt{iban}, \texttt{idcard}, \texttt{passport}, \texttt{socialnumber}, \texttt{taxnum} $\to$ \texttt{account\_number}
  \item \texttt{bankmunicip}, \texttt{bankpostcode}, \texttt{bankstreet}, \texttt{building}, \texttt{city}, \texttt{geocoord}, \texttt{postcode}, \texttt{secaddress}, \texttt{street} $\to$ \texttt{private\_address}
  \item \texttt{cardexpiry}, \texttt{date}, \texttt{dob} $\to$ \texttt{private\_date}
  \item \texttt{email} $\to$ \texttt{private\_email}
  \item \texttt{givenname1}, \texttt{givenname2}, \texttt{lastname1}, \texttt{lastname2}, \texttt{lastname3}, \texttt{title}, \texttt{username} $\to$ \texttt{private\_person}
  \item \texttt{tel} $\to$ \texttt{private\_phone}
  \item \texttt{ip} $\to$ \texttt{private\_url}
  \item \texttt{otp}, \texttt{pass}, \texttt{pin} $\to$ \texttt{secret}
\end{itemize}

We evaluate Privacy Filter accuracy on the remapped PII-Masking-300k dataset using the test holdout only, noting that we did not train on the training portion of the dataset, and we report two sets of metrics described below: baseline and corrected.

When manually reviewing examples that Privacy Filter predicted incorrectly, we found that many spans were either (1) correctly predicted by Privacy Filter but not labeled in the dataset; or (2) labeled in the dataset but correctly detected as background by Privacy Filter. This motivated an experiment where we applied corrections to the dataset labels and obtain a higher confidence estimate of model accuracy.

First, dataset labels were corrected with a conservative post-processing pass after category mapping which aimed to remove false positives in the dataset. The cleanup looked for cases where the same exact substring appeared under multiple active PII categories in a single row, then resolved only high-confidence conflicts. It used deterministic regex and context rules for clear cases like emails, URLs/IPs, phone numbers near phone-related words, dates near date/birth-related words, long numeric IDs, coordinates/postcodes, and placeholder or null values such as \texttt{no disponible}, \texttt{Unknown}, or redaction filler. Placeholder and non-PII values had their labels removed. The baseline metrics are reported on this dataset with only rule-based corrections applied.

While the rule-based approach described above worked well to remove false positives in the dataset, it did not solve the problem of dataset false negatives, such as a \texttt{private\_person} name predicted by Privacy Filter that was not labeled in the dataset. To correct these errors, we used a post-evaluation review step to adjudicate spans detected by the model that are not labeled in the dataset. For each such span, a reasoning model (GPT-5.4 with high reasoning) reviews the full example context along with the detected span and decides whether the span is actually PII, i.e., whether the dataset appears to be missing a valid label. The reasoning model is instructed to classify each span as missing label or undecided if there's not enough information to make a high-confidence decision. Human reviewers then verify those decisions by marking whether they agree or disagree with the adjudication for each disputed span. This process resulted in about 10k spans in 8k examples being marked as missing labels. Human reviewers validated 100 spans and found that the reasoning model's decision agreed with 100\% of their annotations. In the corrected metrics below, we exclude spans identified as missing labels in the adjudication process.

We tried the same approach to adjudicate spans labeled in the dataset but not detected by the model by having a reasoning model examine these spans, but this process reached only 85\% agreement with human reviewers. Therefore we do not make any corrections to the adjudicated dataset false positives.

\subsubsection{Credential Detection in Codebases}
\label{sec:credential-detection-in-codebases-secret-category}

The CredData dataset provides three labels for potential credentials in each example code file: "true positive" (T) / "false positive" (F) / "unknown/other" (X). We map source T labels to our “secret” labels and discard F and X labels. This helps us assess the model's ability to detect credentials accidentally checked into a codebase without falsely triggering on placeholders or high-entropy strings that are not secrets. We assess accuracy on a test split of CredData by counting either tokens or exact spans matched by secret predictions, rather than operating at the line-of-code level. We provide metrics on both a baseline and a corrected version of the dataset, following the same correction procedure outlined in the previous section.

\subsection{Metrics and Matching Rules}
\label{sec:metrics-and-matching-rules}

We report recall, precision, and F1 at the token and span levels, as defined by:

\begin{itemize}[leftmargin=*]
\item \texttt{recall = TP / (TP + FN)}
\item \texttt{precision = TP / (TP + FP)}
\item \texttt{F1 = 2PR / (P + R)}
\end{itemize}

\subsection{Results}
\label{sec:results}

\subsubsection{Benchmark Summary}
\label{sec:benchmark-summary}

We find that Privacy Filter achieves strong overall performance on PII detection and strong recall on secret-credential detection tasks. Note that span-level scores for CredData are lower than token-level scores and are affected by discrepancies in exact span boundaries for spans that are otherwise partially detected. \Cref{tab:benchmarks} shows our performance on both the benchmark datasets as provided and the corrected versions we tested.

\begin{table}[!htbp]
\caption{Privacy Filter performance on benchmark datasets.}
\label{tab:benchmarks}
\centering
\footnotesize
\begin{tabular}{llrrrr}
\toprule
Benchmark Dataset & Scope & Precision (tokens) & Recall (tokens) & F1 (tokens) & F1 (spans) \\
\midrule
PII-Masking-300k (baseline) & All labels & 0.940 & 0.980 & 0.960 & 0.926 \\
PII-Masking-300k (corrected) & All labels & 0.968 & 0.981 & 0.974 & 0.942 \\
CredData (baseline) & \texttt{secret} & 0.747 & 0.965 & 0.842 & 0.617 \\
CredData (corrected) & \texttt{secret} & 0.750 & 0.965 & 0.844 & 0.624 \\
\bottomrule
\end{tabular}
\end{table}

\FloatBarrier
\subsubsection{Fine-tuning Efficiency}
\label{sec:fine-tuning-efficiency}

Users may choose to adapt the labeling policies to their application and tune Privacy Filter to their data domain. We illustrate this point using the SPY dataset (\href{https://huggingface.co/datasets/mks-logic/SPY}{link}), a synthetic collection of legal questions and medical consultations that is slightly out of distribution for Privacy Filter. This dataset labels only information about the question author as PII, excluding other persons referenced in the text (affecting precision), and contains idiosyncratic types of account numbers (affecting recall). We find that fine-tuning on even a small fraction of the training split yields large improvements in validation scores. Training on 10\% of the dataset is enough to drive F1 scores above 96\% and nearly saturates the benchmark. See \Cref{tab:fine-tuning} for performance by training fraction.

As part of the above analysis, we mapped labeled spans in the SPY dataset to the categories supported by Privacy Filter, excluding Privacy Filter categories that have no match in SPY, namely \texttt{secret} and \texttt{private\_date}. See \Cref{tab:remapping} for our complete map.

\begin{table}[!htbp]
\caption{Fine-tuning results on fractions of the SPY training dataset.}
\label{tab:fine-tuning}
\centering
\footnotesize
\begin{tabular}{rrrrr}
\toprule
Train Fraction & Best Epoch & Precision (tokens) & Recall (tokens) & F1 (tokens) \\
\midrule
0\% & 0 & 0.414 & 0.795 & 0.545 \\
1\% & 13 & 0.888 & 0.871 & 0.879 \\
10\% & 39 & 0.963 & 0.960 & 0.962 \\
50\% & 18 & 0.983 & 0.982 & 0.983 \\
\bottomrule
\end{tabular}
\end{table}

\begin{table}[!htbp]
\centering
\caption{Mapping from SPY lables to Privacy Filter labels.}
\label{tab:remapping}
\footnotesize
\begin{tabular}{@{}p{0.1\linewidth}c p{0.2\linewidth}@{}}
\toprule
\texttt{ID\_NUM} & $\to$ & \texttt{account\_number} \\
\texttt{ADDRESS} & $\to$ & \texttt{private\_address} \\
\texttt{EMAIL} & $\to$ & \texttt{private\_email} \\
\texttt{NAME}, \texttt{USERNAME} & $\to$ & \texttt{private\_person} \\
\texttt{PHONE\_NUM} & $\to$ & \texttt{private\_phone} \\
\texttt{URL} & $\to$ & \texttt{private\_url} \\
\bottomrule
\end{tabular}
\end{table}

The Privacy Filter CLI offers utilities for further fine-tuning to a target distribution which enables adapting the predictions to custom decision criteria. Internally, for example, we have tuned a version of the model to a different label set that distinguishes between private and public information of a given type (e.g. \texttt{private\_address} vs. \texttt{public\_address}, such as an official state residence).

\FloatBarrier
\subsection{Stress-Testing OPF Capabilities on Synthetic Data}
\label{sec:stress-testing-opf-capabilities-on-synthetic-data}

We aim to provide users with an understanding of the capabilities and limitations of Privacy Filter. For this purpose, we constructed synthetic toy data sets by prompting frontier GPT models to generate synthetic sequences with PII. Importantly, we did not train on data generated in the same manner.

\subsubsection{Influence of the Category Clue Position}
\label{sec:influence-of-the-category-clue-position}

We wanted to measure how much the model relies on contextual information. We created three separate English-only evaluations for this purpose: 1,181 examples where the category clue is immediately before the PII value, 1,179 examples where the category clue is immediately after the PII value, and 1,200 PII-only examples where clues are omitted entirely. See \Cref{tab:with-without-clues} for a breakdown of how the inclusion and placement of these clues affect performance. \Cref{tab:pii-with-clues} shows three examples that highlight the clue placement or omission in these datasets.

\begin{table}[!htbp]
\centering
\caption{Examples of PII with and without contextual clues.}
\label{tab:pii-with-clues}
\begin{tabular}{rl}
\toprule
\textbf{Prefix:} & This is my home landline number: 442 222 47571 \\
\textbf{Suffix:} & 442 222 47571 is my home landline number. \\
\textbf{Omitted:} & 442 222 47571 \\
\bottomrule
\end{tabular}
\end{table}

All these examples include the same personal identifier but the clue for the PII category is either prefixed, suffixed or omitted entirely. The sentence structure here is always simple and the clue, if provided, unambiguously identifies the personal identifier.

\begin{table}[!htbp]
\centering
\footnotesize
\caption{Privacy Filter performance by PII category, with and without clues.}
\label{tab:with-without-clues}
\begin{tabular}{lrrrrrr}
\toprule
& \multicolumn{3}{c}{Precision} & \multicolumn{3}{c}{Recall}   \\
Category & PII only & Clue $+$ PII & PII $+$ Clue & PII only & Clue $+$ PII & PII $+$ Clue \\
\midrule
\texttt{account\_number} & 0.750 & 0.894 & 0.737 & 0.797 & 1.000 & 0.916 \\
\texttt{private\_address} & 0.821 & 0.989 & 0.940 & 0.214 & 0.414 & 0.368 \\
\texttt{private\_date} & 1.000 & 1.000 & 1.000 & 0.220 & 0.894 & 0.496 \\
\texttt{private\_email} & 0.969 & 0.975 & 0.974 & 0.962 & 1.000 & 0.974 \\
\texttt{private\_person} & 0.782 & 0.912 & 0.793 & 0.885 & 0.994 & 0.955 \\
\texttt{private\_phone} & 0.922 & 0.993 & 0.950 & 0.318 & 0.919 & 0.574 \\
\texttt{private\_url} & 0.618 & 0.855 & 0.730 & 0.835 & 0.942 & 0.872 \\
\texttt{secret} & 1.000 & 0.987 & 1.000 & 0.361 & 0.708 & 0.416 \\
\midrule
Overall & 0.816 & 0.951 & 0.869 & 0.584 & 0.863 & 0.705 \\
\bottomrule
\end{tabular}
\end{table}

\FloatBarrier
\subsubsection{Multilingual Evals}
\label{sec:multilingual-evals}

We evaluate multilingual performance in two complementary ways. First, \Cref{tab:pii-masking-by-language} reports results on PII-Masking-300k broken down by language, which helps us measure how consistently Privacy Filter performs across the multilingual benchmark distribution itself. This view is useful for identifying whether aggregate scores conceal substantial variation across languages.

Second, to study generalization beyond the languages included in PII-Masking-300K data, we constructed additional synthetic multilingual evaluations with roughly the same format. These evaluations extend coverage to languages that were less represented, or not represented, in our synthetic training pipeline and therefore provide a more direct test of out-of-distribution multilingual robustness. \Cref{tab:synthetic-multilingual} reports performance on this expanded set of languages.

Taken together, these two analyses help distinguish in-distribution multilingual performance from broader cross-lingual generalization. They also make it easier to identify where Privacy Filter benefits from multilingual pretraining priors and where performance still degrades because of script, formatting, or language-specific conventions.

\begin{table}[!htbp]
\centering
\caption{Privacy Filter performance on PII-Masking-300K data broken down by language.}
\label{tab:pii-masking-by-language}
\begin{tabular}{lrrrrrr}
\toprule
Language & Number of Examples & Recall & Precision & F1 \\
\midrule
Dutch & 7,457 & 0.937 & 0.892 & 0.914 \\
English & 7,946 & 0.965 & 0.905 & 0.934 \\
French & 8,413 & 0.969 & 0.889 & 0.927 \\
German & 8,120 & 0.965 & 0.890 & 0.926 \\
Italian & 7,976 & 0.959 & 0.886 & 0.921 \\
Spanish & 7,816 & 0.968 & 0.901 & 0.933 \\
\bottomrule
\end{tabular}
\end{table}

\begin{table}[!htbp]
\centering
\caption{Privacy Filter performance on multilingual synthetic data broken down by language.}
\label{tab:synthetic-multilingual}
\begin{tabular}{lrrrr}
\toprule
Language & Number of Examples & Recall & Precision & F1 \\
\midrule
Bengali & 953 & 0.875 & 0.851 & 0.863 \\
Hausa & 941 & 0.801 & 0.720 & 0.758 \\
Hindi & 968 & 0.889 & 0.882 & 0.886 \\
Indonesian & 957 & 0.897 & 0.877 & 0.887 \\
Japanese & 968 & 0.866 & 0.897 & 0.881 \\
Korean & 962 & 0.887 & 0.903 & 0.895 \\
Mandarin Chinese & 971 & 0.921 & 0.913 & 0.917 \\
Modern Standard Arabic & 961 & 0.856 & 0.902 & 0.878 \\
Portuguese & 958 & 0.931 & 0.935 & 0.933 \\
Russian & 957 & 0.890 & 0.900 & 0.895 \\
Turkish & 956 & 0.883 & 0.852 & 0.867 \\
Urdu & 949 & 0.877 & 0.880 & 0.878 \\
Western Punjabi & 781 & 0.853 & 0.847 & 0.850 \\
\bottomrule
\end{tabular}
\end{table}

For the data with the category clue before the PII, we inspected how we perform on languages that are not represented in PII-masking-300k (including English as a baseline), see \Cref{tab:clue-before-multilingual}.

\begin{table}[!htbp]
\centering
\caption{Privacy Filter performance on category-clue-before-PII multilingual data broken down by language.}
\label{tab:clue-before-multilingual}
\begin{tabular}{lrrr}
\toprule
Language & Number of Examples & Precision & Recall \\
\midrule
English & 1,181 & 0.962 & 0.956 \\
Bengali & 1,188 & 0.831 & 0.741 \\
Hindi & 1,193 & 0.887 & 0.790 \\
Indonesian & 1,189 & 0.801 & 0.718 \\
Japanese & 1,189 & 0.900 & 0.758 \\
Mandarin Chinese & 1,191 & 0.926 & 0.786 \\
Modern Standard Arabic & 1,185 & 0.860 & 0.719 \\
Portuguese & 1,192 & 0.890 & 0.760 \\
Russian & 1,193 & 0.914 & 0.793 \\
Turkish & 1,192 & 0.824 & 0.715 \\
Urdu & 1,189 & 0.829 & 0.759 \\
Korean & 1,170 & 0.651 & 0.821 \\
\bottomrule
\end{tabular}
\end{table}

\FloatBarrier
\subsubsection{One-Hop Reasoning Evals}
\label{sec:one-hop-reasoning-evals}

We also evaluate Privacy Filter on examples that require a single step of contextual reasoning to determine that a span should be treated as PII. In these cases, the sensitive value is not explicitly labeled at the point where it appears. Instead, the model must connect it to an earlier statement that defines an alias or reference, such as a phrase indicating that a later token sequence corresponds to an account number or national identification number.

Representative examples include prompts of the form: ``For verification purposes, when I say `marigold' later on, I'm referring to my residential electric utility account number; I'll provide it at the very end,'' followed much later by ``As mentioned earlier, `marigold' is 7281-0543-98217.'' Similarly, another example defines ``starlight'' as a government-issued national identification number before revealing the corresponding value only after a long stretch of unrelated chat context.

\begin{figure}[!htbp]
  \centering
  \includegraphics[width=0.8\textwidth]{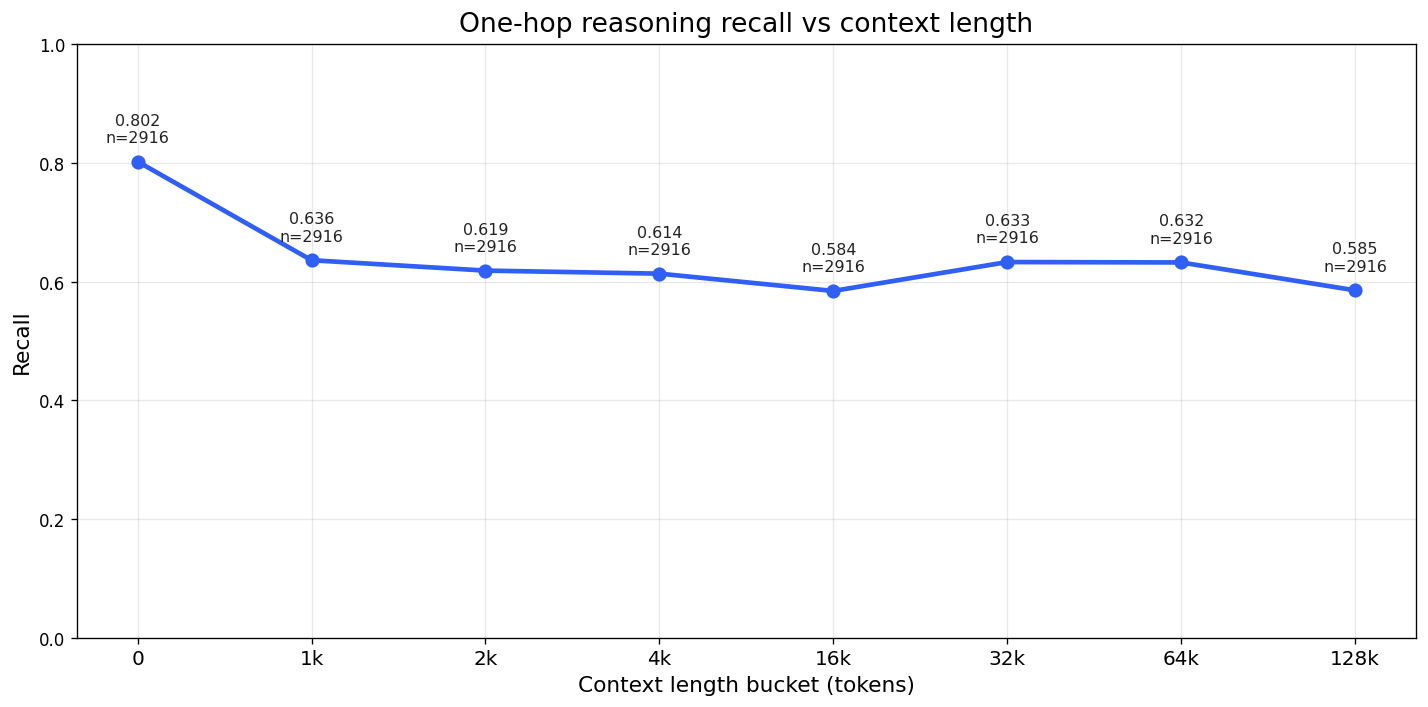}
  \caption{One hop reasoning results. Recall drops as the distance between the alias and its definition grows within the context window.}
  \label{fig:one-hop}
\end{figure}

This evaluation, plotted in \Cref{fig:one-hop}, shows that one-hop reasoning remains difficult for the Privacy Filter. Recall degrades as the relevant contextual cue becomes more distant from the sensitive span, suggesting that Privacy Filter has limited ability to preserve and apply this kind of deferred reference over long context horizons. A likely reason is that such long-range alias-resolution behavior is underrepresented in the current training data. 

\FloatBarrier
\subsubsection{Adversarial Evals}
\label{sec:adversarial-evals}

\begin{figure}[!htbp]
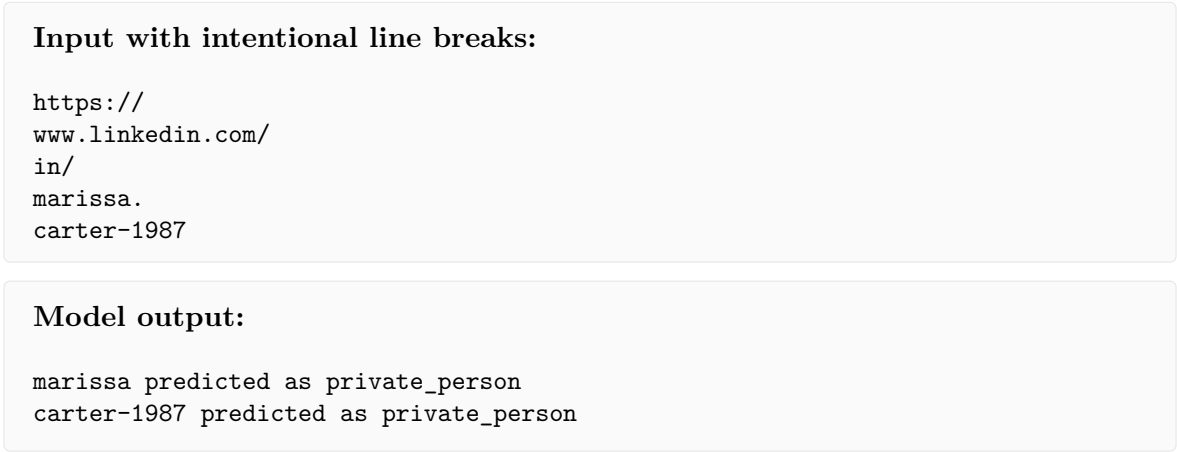

\centering
\begin{tcolorbox}[
  enhanced,
  width=0.94\linewidth,
  colback=gray!4,
  colframe=gray!20,
  boxrule=0.4pt,
  arc=2pt,
  left=8pt,
  right=8pt,
  top=6pt,
  bottom=6pt,
]
\textbf{Input with intentional line breaks:}\par
\begin{Verbatim}[fontsize=\small]
https://
www.linkedin.com/
in/
marissa.
carter-1987
\end{Verbatim}
\end{tcolorbox}

\begin{tcolorbox}[
  enhanced,
  width=0.94\linewidth,
  colback=gray!4,
  colframe=gray!20,
  boxrule=0.4pt,
  arc=2pt,
  left=8pt,
  right=8pt,
  top=6pt,
  bottom=6pt,
]
\textbf{Model output:}\par
\begin{Verbatim}[fontsize=\small]
marissa predicted as private_person
carter-1987 predicted as private_person
\end{Verbatim}
\end{tcolorbox}
\caption{Example showing how explicit line breaks in a URL can lead to fragmented false-positive \texttt{private\_person} predictions. This whole span should be classified as \texttt{private\_url}, but the line breaks lead to error.}
\label{fig:line-break-false-positives}
\end{figure}

In order to understand how Privacy Filter performs in adversarial settings that we might encounter in real data, we created various adversarial evaluations. We start with \textit{confusion category evals}, in which we use frontier models to synthesize sentences where the PII is linked to a non-PII type category but it is still clearly visible from the context that the value might be personal information (examples in \Cref{tab:adversarial-context}). Next, we consider \textit{confusing format alternatives}. For this, we compiled a list of formats that might make the identification of PII harder and generated data according to these various formats, examples in \Cref{tab:adversarial-formatting-alternatives}. Our results are shown in Table~\ref{tab:adversarial-formatting-results}.
Note that the precision for phonetic alphabet can be explained by many name-like tokens in the phonetic spelling (i.e. charlie or oscar). 
Similarly for line breaks, PII such as URLs are split into lines that could incorrectly be identified as \texttt{private\_person}, see \Cref{fig:line-break-false-positives}.

\FloatBarrier

\begin{table}[!htbp]
\centering
\footnotesize
\caption{Examples of adversarial context.}
\label{tab:adversarial-context}
\begin{tabular}{p{0.2\textwidth}p{0.4\textwidth}p{0.39\textwidth}}
\toprule
\texttt{account\_number} & {\ttfamily This is a dice-roll sequence: 298212770451.} & A customer's gift card primary account number \\
\texttt{private\_email} & {\ttfamily For sending me the rehearsal schedule, use my personal email address below. This is a painting title: emily.rose19@icloud.com} & An email address. \\
\texttt{private\_date} & {\ttfamily This is a cheese-aging batch code: 07.08.2001 13:4O.} & A real date and time.\\
\bottomrule
\end{tabular}
\end{table}

\begin{table}[!htbp]
\centering
\caption{Adversarial formatting alternatives used in evaluation.}
\label{tab:adversarial-formatting-alternatives}
\scriptsize
\renewcommand{\arraystretch}{0.92}
\begin{tabular}{p{0.25\textwidth}p{0.25\textwidth}p{0.45\textwidth}}
\toprule
Formatting Alternative & Description & Example \\
\midrule
\texttt{digit\_words} & Digits are spelled out as words. & {\ttfamily If you need to reach me directly, my personal cell number is two six eight - seven two two - one zero four nine.} \\
\texttt{spacing} & Extra or unusual spacing splits chunks or characters. & {\ttfamily This is the compromised API access token from yesterday's deploy, tok\_ 7Qz .P9x 2Kj 8dL 3wV -QeR 1uT .ZxY 4bN \_oP F, so rotate and invalidate it immediately.} \\
\texttt{line\_breaks} & The PII spans explicit newline breaks. & {\ttfamily You can reach Emily directly at her personal email address: emily. carter @proton mail.com for scheduling the home visit.} \\
\texttt{symbol\_substitution} & Some characters or separators are swapped for visually similar non-ASCII characters or emoji. & {\ttfamily Please list the package recipient's full name as Sophia JOhnson.} \\
\texttt{emoji\_word\_replacement} & One word-sized semantic chunk is replaced with a matching emoji. & {\ttfamily For the lease paperwork, my home address is 482 Maple [road], Apt 3B, Brookside, OR 97401\textbackslash{}.} \\
\texttt{at\_dot\_obfuscation} & Email or URL separators are replaced with worded or bracketed forms like [dot]. & {\ttfamily Check out Emily Ross's personal website for her essays and travel photos at www[dot]emily-ross-journal[dot]com.} \\
\texttt{phonetic\_alphabet} & Alphabetic chunks are rendered with phonetic alphabet words. & {\ttfamily My email is charlie.oscar.lima.echo at golf mike alpha india lima dot charlie oscar mike} \\
\texttt{minimal\_name\_fragment} & A person name is unusually short or clipped, with context making it a real name. & {\ttfamily On the clinic intake form, the patient's legal name was recorded exactly as J. Lin, as shown on their insurance card.} \\
\texttt{emoji\_symbol\_replacement} & One symbol-sized part of a name or handle is replaced with a fitting emoji. & {\ttfamily If you're trying to reach me about the carpool, ping my handle luna[moon]beam on the PTA message board.} \\
\bottomrule
\end{tabular}
\end{table}

\begin{table}[!htbp]
\centering
\caption{Privacy Filter performance on adversarial formatting variants.}
\label{tab:adversarial-formatting-results}
\scriptsize
\renewcommand{\arraystretch}{0.92}
\begin{tabular}{p{0.34\textwidth}rrr}
\toprule
Adversarial Axis & Number of Examples & Precision & Recall \\
\midrule
\texttt{confusion\_category\_eval\_overall} & 178 & 0.728 & 0.687 \\
\texttt{spacing} & 270 & 0.699 & 0.631 \\
\texttt{symbol\_substitution} & 247 & 0.824 & 0.907 \\
\texttt{line\_breaks} & 219 & 0.453 & 0.812 \\
\texttt{digit\_words} & 156 & 0.715 & 0.674 \\
\texttt{phonetic\_alphabet} & 115 & 0.273 & 0.694 \\
\texttt{at\_dot\_obfuscation} & 85 & 0.561 & 0.976 \\
\texttt{emoji\_word\_replacement} & 46 & 0.949 & 0.962 \\
\texttt{emoji\_symbol\_replacement} & 31 & 0.906 & 0.935 \\
\texttt{minimal\_name\_fragment} & 24 & 0.885 & 0.958 \\
\bottomrule
\end{tabular}
\end{table}

\FloatBarrier
\section{Conclusion}
\label{sec:conclusion}

Privacy Filter brings strong PII detection to a compact, efficient model that is practical to inspect, adapt, and deploy. Across benchmark and multilingual evaluations, it shows that a small bidirectional classifier can deliver high-quality privacy filtering with low overhead while remaining easy to tune for specific applications.

More broadly, Privacy Filter reflects our view that privacy-preserving infrastructure should be lightweight, inspectable, and usable as part of real production systems. We expect it to be most useful as a flexible component within larger privacy workflows, where its long context, configurable decoding, and fine-tuning capability can support a range of deployment needs. We are releasing this preview to invite feedback from the research and privacy communities and to guide further improvements in model quality, multilingual coverage, and practical deployment.

\section*{Acknowledgements}

We thank the many colleagues across OpenAI and the external teams who contributed to the evaluation, integration, and deployment of Privacy Filter. Their expertise, feedback, and operational support helped bring the model into real-world use.

\end{document}